\documentclass[sigconf]{acmart}

\usepackage{mdframed}
\usepackage{balance}
\usepackage{graphicx}
\usepackage{amsfonts}
\usepackage{amsmath}

\usepackage{amssymb}
\usepackage{algorithm}
\usepackage[noend]{algpseudocode}
\usepackage[utf8]{inputenc}
\usepackage{circledsteps}
\usepackage{comment}
\usepackage[many]{tcolorbox}
\usepackage{multirow}
\usepackage{tikz}
\usepackage{circledsteps}
\usepackage{lscape}
\usepackage{xcolor}
\usepackage{colortbl}
\usepackage{booktabs}
\usepackage{tabularx}
\newcolumntype{C}{>{\centering\arraybackslash}X}
\newcolumntype{L}[1]{>{\raggedright\let\newline\\\arraybackslash\hspace{0pt}}m{#1}}
\newcolumntype{R}[1]{>{\raggedleft\let\newline\\\arraybackslash\hspace{0pt}}m{#1}}
\usepackage{array}
\usepackage{url}
\usepackage{tablefootnote}
\usepackage{hyperref}
\usepackage{adjustbox}
\usepackage{newfloat}

\DeclareFloatingEnvironment[
fileext=loc,
listname={List of examples},
name=Example,
placement=pbth,
]{example}
\usepackage[normalem]{ulem}

\newtcolorbox{box1}{
    enhanced,
    boxrule = 0pt,
    colback = lightgray,
    borderline west = {1pt}{0pt}{main}, 
    borderline west = {0.75pt}{2pt}{main}, 
}

\newtcolorbox{box2}{
    sharpish corners, 
    boxrule = 0pt,
    toprule = 4.5pt, 
    enhanced,
    fuzzy shadow = {0pt}{-2pt}{-0.5pt}{0.5pt}{black!35} 
}

\AtBeginDocument{%
  \providecommand\BibTeX{{%
    \normalfont B\kern-0.5em{\scshape i\kern-0.25em b}\kern-0.8em\TeX}}}

\setcopyright{acmlicensed}
\copyrightyear{2026}
\acmYear{2026}
\acmDOI{XXXXXXX.XXXXXXX}

\acmConference[TechDebt'26]{International Conference on Technical Debt}{April 12--18,
  2026}{Rio De Janeiro, Brazil}

\copyrightyear{2026}
\acmYear{2026}
\setcopyright{cc}
\setcctype{by}
\acmConference[TechDebt '26]{International Conference on Technical Debt}{April 12--13, 2026}{Rio de Janeiro, Brazil}
\acmBooktitle{International Conference on Technical Debt (TechDebt '26), April 12--13, 2026, Rio de Janeiro, Brazil}
\acmPrice{}
\acmDOI{10.1145/3794915.3795778}
\acmISBN{979-8-4007-2485-5/2026/04}

\begin{document}

\title{Investigating Technical Debt Types, Issues, and Solutions in Serverless Computing}

\author{Hasini Sumalee Perera}
\affiliation{%
  \institution{University of Saskatchewan}
  \country{Canada}
}
\email{bhn169@mail.usask.ca}

\author{Zadia Codabux}
\affiliation{%
  \institution{University of Saskatchewan}
  \country{Canada}
}
\email{zadiacodabux@ieee.org}

\author{Fabio Palomba}
\affiliation{%
  \institution{University of Salerno}
  \country{Italy}
}
\email{fpalomba@unisa.it}

\renewcommand{\shortauthors}{Perera et al.}

\begin{abstract}Serverless computing is a cloud execution model where developers run code, and the server management is handled by the cloud provider. Serverless computing is increasingly gaining popularity as more systems adopt it to enhance scalability and reduce operational costs. While it has numerous benefits, it also embodies unique challenges inherent to serverless computing. One such challenge is Technical Debt (TD), which is exacerbated by the complexities of the serverless paradigm. While prior work has investigated the activities and bad practices that lead to TD in serverless computing, there remains a gap in understanding how TD manifests, the challenges it poses, and the solutions proposed to address TD issues in serverless systems. This study aims to investigate TD in the serverless context using Stack Overflow (SO) as a knowledge base. We collected 78,867 serverless questions on SO and labeled them as TD or non-TD using deep learning. Moreover, we conducted an in-depth analysis to identify types of TD in serverless settings, associated issues, and proposed solutions. We found that 37\% of the serverless questions on SO are TD-related. We also identified six serverless-specific issues. Our research highlights the need for tools that can effectively detect TD in serverless applications.
\end{abstract}



\maketitle

\section{Introduction}

Serverless computing is a cloud service model in which the service provider controls the provisioning and management of servers. This architecture allows code to be executed without direct control over underlying resources~\cite{Cinque}. The most prevalent type of serverless computing is Function-as-a-Service (FaaS), which uses platforms such as AWS Lambda, Azure Functions, and Google Cloud Functions. As software architecture patterns have evolved from monolithic architectures to microservices, and then to serverless functions, organizations seek improved maintainability and adaptability for their systems~\cite{Taibi}. Even though adopting serverless computing offers benefits in scalability and efficiency, it can magnify long-term challenges in system quality and performance.

Considering the similarities between serverless computing and microservices, it is reasonable to anticipate that maintainability issues in serverless computing will be significant and ultimately lead to Technical Debt (TD), as is the case with microservices~\cite{Lenarduzzi}. TD arises from the compromises developers make to expedite development. TD may not be an issue right away, but can degrade the system's quality and performance over time, if not managed~\cite{Brown}. In serverless computing, TD can accrue due to rushed architectural decisions, inadequate knowledge of services, minimal process automation, and neglected security practices~\cite{Lenarduzzi, Taibi}. When developers encounter TD issues when implementing software, they frequently turn to community-driven technical Q\&A forums like Stack Overflow (SO) to seek advice, share experiences, and discover solutions. SO hosts 23 million questions, 34 million answers, and 19 million users, and is used by 30\% of software practitioners every day, making it an invaluable knowledge resource for software engineering researchers~\cite{tanzil2025systematic}. These publicly available discussions provide a rich source of data for understanding how TD manifests in real-world serverless applications and how practitioners attempt to resolve it.

Several studies examine the concept of serverless computing, the difficulties developers encounter when implementing serverless applications, and the bad practices that contribute to TD in serverless computing ~\cite{baldini,Leitner,Taibi,wen2021empirical}. Although existing research on serverless computing provides valuable insights, it primarily focuses on the benefits, drawbacks, and challenges of implementing serverless applications.  As far as we are aware, there has been only one study~\cite{Lenarduzzi} that focuses on TD in serverless computing, which was conducted by interviewing three experts and is limited to code, testing, and architectural debt. There is a lack of empirical evidence on how different TD types appear in real-world serverless applications and how they are addressed. Our research aims to bridge this gap by providing a comprehensive analysis of the types, subtypes, and solutions to TD questions in serverless computing, using SO as a knowledge base. This can help provide guidelines to developers on how to minimize TD during the implementation of serverless applications. Moreover, it can help design more effective tools or strategies on how TD issues can be detected and managed in these rapidly evolving environments.

Our study systematically analyzes discussions on SO about TD in serverless computing. We quantify the \emph{prevalence} of TD and \emph{categorize} its types within serverless architectures. Furthermore, we \emph{identify TD issues} exclusive to serverless computing and evaluate the \emph{proposed solutions} discussed by developers. 

The contributions of our study are as follows: 

\begin{enumerate}
    \item The most common TD types in the serverless domain.
    \item A catalog of TD types in the serverless domain.
    \item A catalog of solution types for serverless TD questions.
    \item A comprehensive replication package\footnote{https://doi.org/10.5281/zenodo.17381566}.
\end{enumerate}

\section{Related Work}
\label{sec:rw}

\textbf{Investigating TD using SO. } SO serves as a valuable resource for exploring TD due to its extensive community of practitioners who discuss the questions and challenges they encounter while implementing a system or acquiring new knowledge in specific topics. Gama et al.~\cite{Gama} utilized data mining techniques to pinpoint discussions related to TD on Stack Overflow (SO). After selecting 195 significant discussions from a substantial pool, each discussion was analyzed to identify the types of TD, as well as the activities, strategies, and tools employed in their management. Costa et al.~\cite{Costa} also employed data mining techniques on TD-related discussions from SO to extract empirical findings. Through an analysis process, they identified indicators and forms of debt, along with the quality attributes linked to the occurrence of TD items mentioned in these conversations. Gama et al.~\cite{Gama2} investigated software engineering practices from a developer's perspective, focusing on how developers typically identify TD items in their projects. They analyzed 140 TD-related discussions on SO, demonstrating that practitioners on SO commonly discuss identifying TD and identified 29 distinct low-level indicators across code, infrastructure, architecture, and tests. They subsequently categorized the indicators into three groups based on the type of debt they were intended to identify, including low- and high-level indicators. These studies provide a broader examination of TD on SO to analyze different aspects of TD. 

Kozanidis et al.~\cite{Kozanidis} employed both automated and manual approaches to identify the questions related to TD. Questions extracted from SO were carefully evaluated to perform quantitative and qualitative analyses across various aspects, including TD types, question length, urgency, sentiment, and themes. They used machine learning to classify SO questions as TD or non-TD. Their research explored the possibility of using machine learning to identify TD and its types within SO questions.  While their study investigates TD in general, we focus on identifying TD in serverless-related questions on SO. Edbert et al.~\cite{Edbert} collected data from SO to explore the TD in security-related questions on SO. They extracted data from predefined security tags in Yang et al.~\cite{Yang}, classified the extracted security questions as TD or non-TD, and explored the characteristics of security-related TD questions on SO. We followed a similar approach to their study, but we focused on investigating the issues and proposed solutions of serverless-related TD questions on SO. Titan et al.~\cite{Titan} used 14 terms related to architecture smells to extract relevant questions on SO, resulting in the collection of 207 questions. They utilized the grounded theory method to analyze these questions, focusing on different aspects of architectural smells. This paper examines discussions obtained through specific terms related to architectural smells.

\textbf{Studies on Serverless Computing.} Taibi et al.~\cite{Taibi} interviewed experienced software engineers proficient in serverless technology to identify bad practices in serverless and propose solutions to address them. This process identified six bad practices related to applications developed using serverless architecture. Leitner et al.~\cite{Leitner} conducted a mixed-method study to investigate serverless technologies in industry. Their findings included prevalent design patterns in serverless, as well as the perceived advantages and difficulties developers encounter when utilizing serverless computing. Wen et al.~\cite{wen2021empirical} conducted empirical research in which they manually evaluated 619 examples and analyzed 22,731 SO questions related to serverless computing to develop a taxonomy of serverless challenges. Baldini et al.~\cite{baldini} presented a programming paradigm along with additional relevant features of serverless computing. Additionally, they identified the types of application workloads that are suitable for serverless computing platforms.

\textbf{TD in Serverless.} Exploring TD in serverless applications is a relatively understudied topic. Lenarduzzi et al.~\cite{Lenarduzzi} interviewed three experts to determine the activities that contribute to the accumulation of TD across code, test, and architecture debt. They investigated how TD is accrued in serverless functions, identifying a preliminary set of issues that are indicators of TD. Their analysis shed light on the types of TDs most likely to affect serverless computing.

\textbf{Studies on Microservices.} Contrary to serverless computing, Microservices, the predecessor of serverless computing, was more extensively studied. Despite their similarities,  their main difference is in their roles. Microservices is a design technique for applications, whereas serverless functions are the architecture for running parts or entire applications. In particular, microservices can be hosted within a serverless architecture. Toledo et al.~\cite{Toledo} interviewed 25 practitioners working with microservices in seven large companies. They identified 16 architectural TD issues, their negative impacts, and common solutions to repay each debt and its related costs. Pigazzini et al.~\cite{Pigazzini} studied a real-life project involving approximately 1,000 services within a large international company, quantitatively analyzing the documentation and conducting interviews. The research studies discussed above mainly focus on investigating TD in microservices through expert interviews. On the contrary, our emphasis is on serverless architecture, rather than microservices, and on analyzing TD through the mining of SO questions.

\textbf{Summary.} Several studies have explored various aspects of TD using SO. \textbf{Our study exclusively focuses on TD within the context of questions on serverless on SO}. Moreover, in previous work, expert knowledge was used to gain insights into TD in serverless architectures, with a focus on three debt types (code, test, and architecture debt), \textbf{whereas our study goes a step further by exploring additional TD types using SO as the knowledge base}. These two key differences allow our study to provide evidence-based insights into the prevalence of TD, serverless-specific challenges encountered by the community, and the availability of solutions. In addition, architectures such as microservices have received extensive research on different types of smells and debt, and on how they are manifested. In our study, we aim to understand how TD manifests in serverless systems. These findings may inform research and practice by highlighting common issues, identifying gaps in existing solutions, and guiding the development of effective tools and practices to address TD in serverless computing.

\section{Methodology}
\label{sec:method}

\begin{figure*}[h]
\centering
    \includegraphics[width=\textwidth]{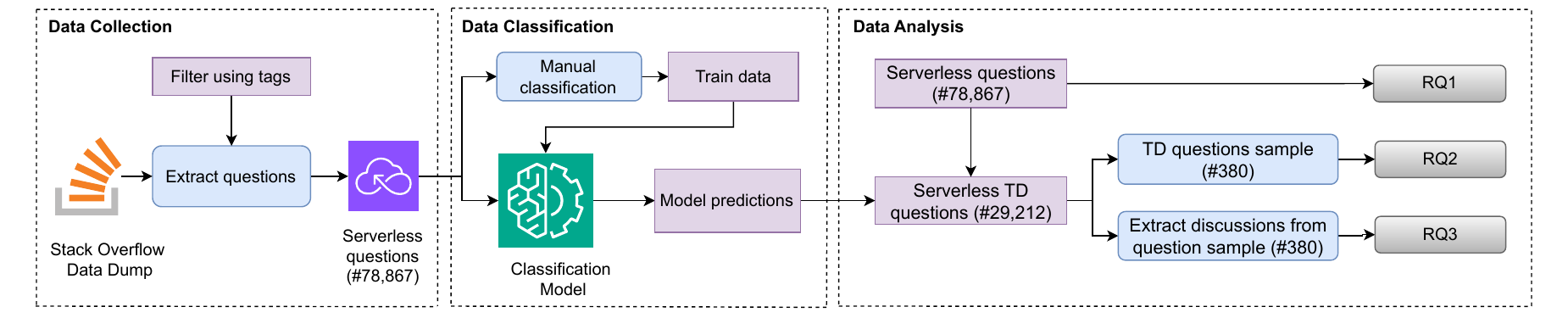}
    \caption{Study Design}
\label{figure:methodology}
\end{figure*}

\subsection{Goal \& Research Questions}
\label{subsec:methodA}

The \emph{goal} of this study is to investigate the prevalence, types, specific issues, and solutions of serverless-related questions on SO, with the \emph{purpose} of characterizing TD in serverless computing and identifying the solutions proposed by practitioners. The \emph{perspective} is of both researchers and practitioners. The former are interested in gaining a deeper understanding of TD in serverless computing, with the aim of identifying requirements that may lead to the development of novel automated approaches better suited to practitioners' needs. The latter is interested in understanding how TD manifests and the strategies other practitioners use to address it, with the aim of improving the quality assurance mechanisms in their own contexts.

Specifically, we pose four Research Questions (RQs):

\textbf{RQ1: To what extent do SO questions in the serverless domain indicate TD?}
\textit{Rationale: }We classify TD questions into various types and subtypes in order to understand which TD types are most often explored on SO. We aim to compute the proportion of TD present in serverless-related questions on SO and investigate how SO users discuss serverless-related queries. This preliminary analysis can provide insight into the common types and subtypes of TD in serverless environments, helping raise awareness of the issue's practical relevance.

\textbf{RQ2: What are the types of TD in serverless computing?}
\textit{Rationale: }We categorize the classified TD questions into different types and subtypes to understand which types of TD are most commonly discussed on SO. The objective is to gain an understanding of the prevailing types of TD discussed by the SO community and identify subtypes specific to each type, which may be instrumental in prioritizing future research and informing practitioners about the types of TD most frequently encountered in practice.

\textbf{RQ3: How is TD in serverless mitigated?}
\textit{Rationale: }We aim to understand how serverless TD questions are addressed, providing insights into the solutions practitioners propose to deal with TD in serverless architectures. We analyze the answers to serverless TD questions to identify the types of answers available. We also analyze the availability of answers and accepted answers (resolved problems) for each TD type. 

Figure \ref{figure:methodology} outlines our methodology, which includes three main phases: data collection, classification, and analysis. Initially, we extracted serverless-related questions from SO and then classified them as TD-related or not. Subsequently, we analyzed the TD-related questions to address our RQs. Our methodology adhered to established guidelines by Wohlin et al.~\cite{Wohlin} and the ACM/SIGSOFT empirical standards\footnote{https://github.com/acmsigsoft/EmpiricalStandard}.

\subsection{Data Collection}
\label{subsec:methodB}

We extracted our data from the publicly available Stack Exchange Data Dump\footnote{https://archive.org/details/stackexchange}, which provided a dataset of all questions on the site from January 2014 to March 2025. Quite noteworthy is the fact that the year 2014 was significant in that it was the year of releasing the concept of serverless computing in an AWS re:Invent’ event\footnote{https://aws.amazon.com/blogs/developer/aws-reinvent-2014-recap-2/}, thus bringing it into the world. Therefore, we have considered a reasonable time period for data collection that accurately reflects the introduction of serverless architecture to the world.

To ensure accurate retrieval of SO questions, we selected the tags most commonly used by the SO community to extract serverless questions. This was done by analyzing existing research on serverless, which used terms that implied the serverless paradigm. Wen et al.~\cite{wen2021empirical} extracted a total of 13 tags as related to serverless computing. These tags included  \emph{serverless-architecture, serverless, serverless-framework, aws-serverless, aws-lambda, aws-sam, aws-sam-cli, serverless-offline, vercel, serverless-plugins, localstack, faas,} and \emph{openwhisk}. We used the same tags and expanded the list based on existing studies. Two additional tags, \emph{azure-functions} and \emph{google-cloud-functions}, were identified as related to the concept of serverless, since a few studies~\cite{Lenarduzzi,Taibi, li2022serverless} reported on these frameworks as serverless computing services delivered by major cloud providers, similar to AWS Lambda. Therefore, we selected 15 tags to extract serverless-related posts. This curation and filtering yielded a substantial dataset comprising 78,867 SO questions on serverless computing.

\subsection{Data Classification}
\label{subsec:methodC}
The data classification stage consisted of two phases: manual classification and automated classification. We first manually classified serverless SO questions into two categories (TD-related and non-TD-related). Next, we used the classified questions as a training dataset for a binary text classification model we built in the next phase. Manual classification was done by two annotators, namely (i) the first author, who has 4 years of software development experience and 2 years of research experience in software engineering, and (ii) a final year PhD student who has 6 years of research experience in software engineering. We evaluated each question against the Dagstuhl TD definition~\cite{Avgeriou} to determine whether each related to TD. In this stage, the annotators individually classified a sample of 150 serverless SO questions as TD-related or non-TD-related.  Cohen's Kappa interrater agreement was 0.83. They discussed and resolved all the disagreements. Then, the annotators continued manually classifying the dataset until they identified 406 TD-related questions. They selected a similar number (419) of non-TD-related questions to create the training dataset. A similar number of non-TD questions were selected to create a balanced dataset for training the binary text classifier. This resulted in a total of 825 SO questions as the training dataset.

We then followed the steps of Kozadinis et al.~\cite{Kozanidis} to develop a binary text classification system. The steps involved: (i) pre-process the collected questions related to the serverless paradigm, (ii) develop a binary text classification model to identify TD questions on the serverless paradigm, (iii) use the classification model to classify the extracted serverless SO questions as related to TD or not.

First, we removed duplicate questions from the dataset. However, we did not find any duplicate questions. Hence, we proceeded to preprocess the questions related to serverless concepts collected from users, ensuring the data was formatted correctly.  We removed punctuation and converted all text to lowercase. Then, we removed SO-related HTML or markdown elements and appearances of source code, as these do not contain any information of lexical or semantic importance.

Subsequently, we trained five models, consisting of four traditional machine learning models used by Kozanidis et al.~\cite{Kozanidis} for the binary text classification of SO questions, Logistic Regression, Random Forest classifier~\cite{Breiman}, Support Vector Machine (SVM), Naive-Bayes classifier, and a deep learning model, BERT~\cite{Devlin}, to identify serverless TD questions and adhere to the guidelines established. BERT was selected because it is an advanced transformer-based model suitable for text classification. Our classification models were evaluated using standard metrics consistent with established practices, namely precision, recall, and the F1-score. During the model training phase, the models achieved the following F1-score: fine-tuned BERT 0.86, Logistic Regression 0.81, Random Forest classifier 0.82, Naive-Bayes classifier  0.81, and SVM 0.78. The model with the highest F1-score was selected to predict the class of the rest of the serverless SO questions, specifically the serverless SO questions collected during the data collection phase. We use the collected serverless SO questions to conduct an in-depth analysis of TD- and non-TD-related questions on SO and to gain insights into various aspects of serverless questions.

\subsection{Data Analysis}
\label{subsec:methodD}

During the final phase of the study, we analyzed the serverless TD questions to identify the prevalence of TD, its types, and how TD is mitigated in SO questions. 

To investigate the prevalence of TD (RQ1), we built a classification model as explained in Section~\ref{subsec:methodC}.
For further analysis of the types of TD in serverless (RQ2), we followed a content analysis approach~\cite{mayring2014qualitative}. There are different taxonomies for TD classification~\cite{Pina, Alves, AlvesMappingStudy, Rios, Li}. We used the TD type taxonomy presented by Li et al.~\cite{Li} to categorize each selected TD question in the sample into ten types: \emph{architecture, build, code, defect, design, documentation, infrastructure, test, requirement,} and \emph{versioning}. We first selected a statistically significant sample (95\% confidence and 5\% margin of error of the 29,212 serverless TD questions), resulting in 380 questions. During the manual categorization process, we identified one additional TD type, \textit{security debt}, in addition to the 10 TD types included in the taxonomy by Li et al.~\cite{Li}. Therefore, in total, we considered 11 types of TD when categorizing the serverless TD question sample (of 380 questions). To reduce bias, a random sample of 50 questions was independently assessed by the PhD student (the second annotator in Section~\ref{subsec:methodC}). We used Cohen's Kappa~\cite{cohen1960coefficient} to calculate the inter-rater agreement. This yielded a value of 0.94, indicating near-perfect agreement among the raters.

We then used open coding~\cite{blair2015reflexive} to identify the subtypes for each TD type. We first broke the question text into smaller parts during the open coding process to identify key information related to a particular serverless computing problem. Then, using the information we extracted from the text, we assigned each question a ``code.'' We followed the same procedure for the subsequent questions. We classified the questions into one of the codes we had already identified; if not, we created a new code. This resulted in 49 codes. One author conducted the analysis by categorizing the SO questions into types and subtypes.

To investigate the proposed solutions to the serverless TD questions (RQ3), we conducted a qualitative analysis of a sample of 380 questions to further identify the problems in serverless TD questions and how they have been addressed. To achieve this, we reviewed the complete discussions of the 380 SO questions, including the answers. We used open coding to categorize the solutions by availability and type. For each TD type, we examine the availability of solutions and the accepted solutions to identify those that require critical attention.

\section{Results}
\label{sec:results}

\subsection{RQ1: To what extent do SO questions in the serverless domain indicate TD?}
\label{subsec:resultsA}

We used the BERT model~\cite{Devlin} to classify the serverless-related questions we extracted from SO as TD- or non-TD-related, which showed the highest F1-score (0.86) among the five models we trained. We used pre-defined weights to train the model and fine-tuned it for our classification task. Our model identified 29,212 of 78,867 serverless-related questions on SO as TD-related, representing 37\% of the serverless questions on SO. This significant proportion highlights the pressing need for additional research in the serverless domain. Nonetheless, it is worth noting that the BERT model's F1-score was 0.86. The validity of the classification results depends on the model's accuracy. Therefore, the percentage presence of TD in serverless may differ slightly from 37\%. To mitigate these threats, we manually classified a statistically significant sample (95\% confidence, 5\% margin of error), resulting in 383 SO questions. We used it to evaluate the model's performance. This resulted in an F1-score of 0.82, which is close to the model's F1-score. 

\begin{tcolorbox}
  \textit{``Anyone know how to set a websocket's content handling strategy to binary in the serverless framework? I have a websocket defined as follows: [code] I'm looking for a sls approach since all of my infra is managed in sls and I don't want to absorb the \textbf{tech debt} migrating (unless I have to)''}
\end{tcolorbox}

\begin{tcolorbox}
  \textit{``Our Google Analytics data events are exported to BigQuery tables. I have reports that need to run when the events data arrives which are set up as AWS lambdas with python code (for various reasons and I can't immediately move these to be Google Cloud Functions etc). Is it possible to have the creation of a table trigger a lambda? At present, I have a lambda periodically checking to see if the table has been created which seems \textbf{suboptimal}. Eventarc looks like it might possibly be the way to monitor for the creation event at the BigQuery end but it doesn't seem obvious how you'd interface with AWS. Any genius ideas? I have dug repeatedly through StackOverflow, but can't see a match for this issue''}
\end{tcolorbox}

As shown in Examples I and II (verbatim), SO users explicitly discuss TD-related queries and use terms such as ``suboptimal'' to refer to TD. In both examples, users are seeking better solutions to the problems they encounter. The results of our classification model indicate that 37\% of serverless questions on SO are related to TD. This provides a brief insight into the prevalence of TD in serverless computing. When SO users ask questions on the platform, it is expected that a certain number of those questions will translate to TD~\cite{Kozanidis, Edbert}. Also, it should be noted that the number of serverless questions that depict TD is significant and needs further investigation.

\begin{box2}
\textbf{RQ1 Summary:} Out of the 78,867 serverless questions on SO, 37\% were identified as TD related questions. SO users discuss TD-related questions explicitly using words such as ``tech debt'' and others that indicate TD, such as ``suboptimal.''
\end{box2} 

\subsection{RQ2: What are the types and subtypes of TD in serverless computing?}
\label{subsec:resultsB}

We followed the taxonomy defined by Li et al.~\cite{Li} to categorize the TD questions into TD types. We manually categorized a sample of 380 questions into 11 TD types.

According to Table~\ref{table:rq2}, the most prevalent type of TD is \textit{code debt}. This is followed by \textit{design debt}, the second-most prevalent type of TD, and \textit{versioning debt}, the third-most prevalent. The least common type of TD in serverless SO questions is \textit{infrastructure debt}. However, we did not find any questions related to \textit{requirement debt}. Therefore, based on RQ2, we identified 10 types of TD and their prevalence in serverless computing. Table~\ref{table:examples-rq2} also includes an example of each type of TD, which can also be found in the replication package$^1$, and the number of TD instances for each type.

We looked further into the problems that occur in serverless TD questions. As depicted in Table~\ref{table:rq2}, we identified the problems associated with each type of TD. We used open coding to categorize the TD types further based on the nature of the issue. This resulted in 49 codes. We identified six subtypes of \textit{architecture debt} issues, with those related to architecture selection being the most common. For \textit{build debt}, deployment-related issues were the most prevalent. \textit{Code debt} encompasses a range of issues, from general implementation concerns to data transfer between functions. We identified four subtypes for \textit{defect debt}. Most of the \textit{design debt} issues were related to function timeouts. Lack of documentation and incorrect documentation were the subtypes of \textit{documentation debt}. \textit{Infrastructure debt} had five subtypes, with scaling being the most common. Authentication and authorization-related issues were the most common among \textit{security debt}. \textit{Test debt} reported issues in testing locally, testing deployed functions, and mocking API calls. The most common subcategory of \textit{versioning debt} relates to upgrading to a new version. To identify issues specific to serverless computing, we compared our results with the existing literature and identified 6 serverless-specific issues. This is further discussed in Section~\ref{subsec:discussionA}.

\begin{table}[!htbp]
\caption{Types and Subtypes of TD}
\centering
\begin{tabular}{L{1.8cm} L{3.7cm} R{0.9cm} R{0.9cm}}
 \toprule 
 \textbf{Type} & \textbf{Sub Type} & \textbf{Count} & \textbf{Total} \\
 \midrule

Architecture & Scaling the application & 4 & 40 \\
& Architecture selection & 11 & \\
& Architecture change & 6 & \\
& Architecture implementation & 9 & \\
& Cold start & 5 & \\
& Cost related & 5 & \\
\midrule

Build & Deployment related & 18 & 28 \\
& Build configurations & 10 & \\
\midrule

Code & Database CRUD operations & 22 & 129 \\
& Function I/O & 1 & \\
& Add/access logs & 6 & \\
& Invoking functions & 19 & \\
& Handling media files & 11 & \\
& Data transfer between services & 21 & \\
& Invoking external APIs & 8 & \\
& Handling events & 4 & \\
& Connection between services & 14 & \\
& Concurrent executions & 8 & \\
& Refactoring & 1 & \\
& Implementation & 14 & \\
\midrule

Defect & Clearing data & 6 & 24 \\
& Query optimization & 4 & \\
& View logs & 2 & \\
& Function specific & 11 & \\
\midrule

Design & Service specific & 12 & 49 \\
& Function timeout & 14 & \\
& Size limitations & 2 & \\
& Design patterns & 5 & \\
& Invoking functions & 8 & \\
& Scheduling tasks & 5 & \\
& Directory structure & 2 & \\
& Shared services & 1 & \\
\midrule

Documentation & Lack of documentation & 15 & 15 \\
\midrule

Infrastructure & Scaling & 4 & 8 \\
& Minimizing cost & 1 & \\
& Building infrastructure & 1 & \\
& Memory limitations & 1 & \\
& Monitoring services & 1 & \\
\midrule

Security & Secrets management & 7 & 31 \\
& Authentication and authorization & 15 & \\
& Accessing other services & 3 & \\
& Limiting access & 6 & \\
\midrule

Test & Testing locally & 3 & 9 \\
& Testing deployed functions & 3 & \\
& Mocking API calls & 3 & \\
\midrule

Versioning & Upgrading the version & 35 & 47 \\
& Downgrading the version & 2 & \\
& Version selection & 2 & \\
& Version capabilities & 4 & \\
& Version deprecation & 4 & \\
\bottomrule
\end{tabular}
\label{table:rq2}
\end{table}

\begin{table*}[!htbp]
\caption{Examples of Types of TD with Distribution}
\centering
\begin{tabular}{L{2.2cm} L{9.8cm} R{2cm} R{2cm}}
 \toprule
 \textbf{Type} & \textbf{Example} & \textbf{Total Count} & \textbf{Percentage (\%)} \\
 \midrule

Architecture &
``We have a microservices architecture developed on Google cloud. Actually the microservices are all running on cloud run and talk each other with rest (sync) or with pub/sub (async).
...We have different services which can safely be updated with a scheduled cron, so every one hour...''%
\tablefootnote{https://stackoverflow.com/questions/63427991} &
40 & 10.5 \\
\midrule

Build &
``...But I repeatedly fail to build successfully respectively to have the Unit tests running successfully in the pipeline. Locally everything works fine. What do I have to do to get the Function App build and unit tested...''%
\tablefootnote{https://stackoverflow.com/questions/58076903} &
28 & 7.4 \\
\midrule

Code &
``...I'm currently refactoring my queries ... Is this the best practice, as the article suggests, in a Serverless Framework API? Lastly, should I instead open the connection my handler and pass pgPool as an argument to my models?...''%
\tablefootnote{https://stackoverflow.com/questions/75950215} &
129 & 33.9 \\
\midrule

Defect &
``I've noticed that this folder has grown up to 1.6Gb of data, it seems to keep there copies of previous releases and I just wonder if I can remove old versions and/or if there's a way to do that automatically.
...''%
\tablefootnote{https://stackoverflow.com/questions/60709896} &
24 & 6.3 \\
\midrule

Design &
``...Is it necessary to follow OOPS concept and create classes in serverless architecture? Can you point me to some sample GIT repo showcasing the framework for developing an object oriented lambda? AWS lambda gets activated on request and dies down when request is served. It is not like server which is active 24hours without request. What is best design pattern or architecture or coding paradigm to follow in AWS lambda code for python language which also supports modular programming?...''%
\tablefootnote{https://stackoverflow.com/questions/69471022} &
49 & 12.9 \\
\midrule

Documentation &
``...Is that possible using Cloud Functions (generate a token based on the Firebase Service Account), if so, how do I obtain it? I could not find any documentation related to generating access tokens within a Cloud Function....''%
\tablefootnote{https://stackoverflow.com/questions/61881693} &
15 & 3.9 \\
\midrule

Infrastructure &
``...AWS Lambda would do the job perfect if it wasn't for the lack of Python 3-support...''%
\tablefootnote{https://stackoverflow.com/questions/38940140} &
8 & 2.1 \\
\midrule

Security &
``...What are the security risks? I know they are used to sign requests...''%
\tablefootnote{https://stackoverflow.com/questions/59653206} &
31 & 8.2 \\
\midrule

Test &
``I'm writing unit tests for a serverless application in TypeScript, and I'd like to mock the AWS SDK. Unfortunately I have not found many existing type definitions for popular AWS mocking projects. In particular I'd like to use the aws-sdk-mock library, but without its type definitions I can't....''%
\tablefootnote{https://stackoverflow.com/questions/50978706} &
9 & 2.4 \\
\midrule

Versioning &
``AWS will soon stop supporting Node 12 and prior versions as it's approaching end of life. I have concerns if I upgrade from node 12 to node 16 if there's going to be any sytnax or semantic changes in the new version of node. I searched the internet and found one case where everything went successful; however, I am not certain. I'm wondering if anyone knows if Node 12 vs Node 16 has any critical changes that could potentially break the code or not be backward compatible?''%
\tablefootnote{https://stackoverflow.com/questions/73860735} &
47 & 12.4 \\

\bottomrule
\end{tabular}
\label{table:examples-rq2}
\end{table*}

\begin{box2}
\textbf{RQ2 Summary:} The most prevalent type of TD in serverless-related questions was \textit{code debt}, with \textit{design debt} and \textit{versioning debt} as the second and third common types of TD, respectively. The least common type of TD was \textit{infrastructure debt}. We identified 49 TD subtypes in serverless SO questions.
\end{box2}

\subsection{RQ3: How is TD in serverless mitigated?}
\label{subsec:resultsD}

To understand how SO users have answered the TD-related questions, we categorized the solutions into four main types: 1) not answered, 2) not possible, but an alternative was suggested, 3) not possible, and 4) a solution was provided. For the questions with provided solutions, we further broke them down into six categories, as shown in Table~\ref{table:rq3}. We categorized the answers to serverless questions, but noticed that the solution categories we identified were not specific to serverless and can be applied to SO questions from other domains. However, we present the results from a serverless perspective. We further investigated the availability of answers and accepted answers for each TD type.

\begin{table}[!htbp]
\caption{Percentage of Questions without Answers \& Accepted Answers}
\label{table:rq32}
\begin{tabular}{L{1.8cm} R{2.9cm} R{2.9cm}}
\toprule
TD Type & Percentage of Questions without Answers & Percentage of Questions without Accepted Answers \\
\midrule
Architecture   & 12.5 & 67.5 \\
Build          & 17.9 & \textbf{75.0} \\
Code           & 12.4 & 51.2 \\
Defect         & 25.0 & 62.5 \\
Design         & 20.4 & 69.4 \\
Documentation  & \textbf{53.3} & 66.6 \\
Infrastructure & 12.5 & 50.0 \\
Security       & 12.9 & 48.4 \\
Test           & 33.3 & 66.6 \\
Versioning     & 14.9 & 41.1 \\
\bottomrule
\end{tabular}
\end{table}

Table~\ref{table:rq32} shows the percentage of questions without answers and the percentage of questions without accepted answers for each type of serverless TD questions. \textit{Documentation debt} has the highest percentage (53.3\%) of unanswered questions, indicating it needs more attention from the SO community. \textit{Code debt} questions have the lowest percentage (12.5\%) of unanswered questions, indicating that the majority are answered on SO. Based on the absence of accepted answers, \textit{build debt} had the highest percentage, with 75\% of questions lacking accepted answers. \textit{Versioning debt} has the lowest percentage (41.1\%) of questions without accepted answers.

\begin{table*}[!htbp]
\caption{Solution of Serverless TD Questions}
\centering
\begin{tabular}{L{3cm} L{3cm} L{9.5cm} R{1cm}}
\toprule 
\textbf{Solution Availability}                  & \textbf{Solution Type}                      & \textbf{Definition} & \textbf{Count} \\ 
\midrule 
\multirow{6}{*}{Solution provided}     & Code modifications                      & Implementation of the solution is provided along with the code snippets.           &   106    \\ \cmidrule(lr){2-4} 
& Alternative solution               & The solution suggested in the question is incorrect. Therefore, an alternative method is suggested.           &  14     \\ \cmidrule(lr){2-4} 
 & Possible suggestions               & Multiple suggestions to approach the problem are suggested as it is unable to find the exact solution with the given information.           & 16      \\ \cmidrule(lr){2-4} 
& Debugging and troubleshooting            & Clear guidelines, including specific instructions to pinpoint and solve the problem, are suggested.           &  13     \\ \cmidrule(lr){2-4} 
  & Conceptual understanding           & The solution is comprised of a detailed explanation of the concept asked in the problem.           & 37      \\ \cmidrule(lr){2-4} 
  & Knowledge resources & Official documentation or implementation of the solution to a similar problem is referenced.           & 84     \\ \cmidrule(lr){2-4} 
  & Tool suggestions & Suggested the use of a tool to solve the problem. & 3 \\
  \cmidrule(lr){2-4} 
  & Best practices & Provided suggestions based on general guidelines or industry standards. & 6 \\ 
\cmidrule(lr){2-4} 
 & Configuration changes & Suggested adjustments to the environment, services, or tools. & 21 \\ 
\midrule
Not possible &   & It is impossible to achieve a solution to the problem due to the limitations of the services.           &  5     \\ 
\midrule
Not possible, alternative given &                                    & It is impossible to achieve a solution, but an alternative approach to solving the problem is suggested.           & 8      \\ 
\midrule
Not answered                           &                                   & A solution has not been provided.           &  67     \\ 
\bottomrule 
\end{tabular}
\label{table:rq3}
\end{table*}

\begin{box2}
\textbf{RQ3 Summary:} The analysis of the types of TD in serverless computing revealed various issues within each category of TD. \textit{Documentation and build debt} were identified as types of TD with the highest percentages of unanswered and unresolved questions. Notably, 78\% of the questions received responses from SO users, while 18\% remained unanswered or posed unsolvable challenges. As potential solutions to the question asked, SO users provided implemented code, detailed explanations, multiple possible solutions, step-by-step guidelines, and references to examples.
\end{box2}

\section{Discussion}
\label{sec:discussion}

\subsection{Serverless Specific Issues}
\label{subsec:discussionA}

In serverless architectures, certain TD issues emerge that are unique to serverless architectures. These issues often stem from challenges in developing serverless applications, particularly in performance, operations, and developer experience~\cite{Van}. In Section~\ref{subsec:resultsB}, we identified types and subtypes of TD that are present in serverless-related SO questions. However, most of these issues (e.g., CRUD database operations, upgrading versions, invoking external APIs, etc.) also apply to other architectures. Therefore, among the TD issues we identified in Section~\ref{subsec:resultsB} below, we elaborate on those specific to serverless. Detailed examples of questions of each TD type can be found in the replication package$^1$.

\textbf{Scaling the Application. }Serverless computing has an auto-scaling feature where the necessary resources are allocated based on the demand~\cite{Tran}. In serverless computing, scaling is handled automatically by the cloud provider. This abstracts away the infrastructure management from the developer~\cite{Lloyd, Somma}. However, this can lead to unexpected challenges. For instance, serverless functions may face limitations in scaling, especially when dependent services or resources do not scale at the same rate. This issue is unique to serverless because, in traditional architectures, scaling typically involves scaling entire server instances, which developers can directly manage and optimize. This finding aligns with the study of Baldini et al.~\cite{baldini}, which identified scaling as an open problem in serverless.

\textbf{Cold Start.} Cold start latency is a well-known issue in serverless computing. This occurs when a function is invoked after an idle period, resulting in a delay. This latency can introduce unexpected delays in function startup, thereby negatively affecting performance~\cite{Yu, Agarwal, Ghorbian, Vahidinia, Verma, Ebrahimi, Karmazadeh}. Unlike traditional architectures, where servers are always running and readily available, serverless functions are stateless and may require initialization before execution, which can lead to cold-start issues. This finding corroborates those of Baldini et al.~\cite{baldini} and Wen et al.~\cite{wen2021empirical}, who identify calling to zero as a key differentiator in serverless, which ultimately leads to cold-start problems.

\textbf{Add/Access Logs. }Logging in serverless computing refers to the logs produced when a serverless function is run. This can be used to monitor and debug the applications. Serverless functions often run in ephemeral environments, making traditional logging practices more challenging. Accessing and aggregating logs from these short-lived instances can be cumbersome, needing specialized tools or services~\cite{Manner}. This contrasts with traditional server-based architectures, where logs are typically stored on persistent disks and can be accessed directly. The ephemeral nature of serverless functions requires a rethinking of logging strategies, often leading to TD when not properly managed. Baldini et al.~\cite{baldini}, Lenarduzzi et al.~\cite{Lenarduzzi}, and Wen et al.~\cite{wen2021empirical} identified similar problems in monitoring and debugging serverless functions, and inadequate or verbose logging, respectively.

\textbf{Minimizing Cost. }Cost optimization is both a benefit and a challenge in serverless computing. The pay-per-use model means that costs can quickly escalate if the functions are not optimized for execution time and resource usage~\cite{Sedefo, Sarroca}. Unlike traditional server-based applications, serverless costs are more dynamic and require constant monitoring and optimization. This can lead to TD if costs are not carefully considered during the design and development phases. Our findings align with those of Baldini et al.~\cite{baldini}, who identified cost as a fundamental challenge in serverless.

\textbf{Secrets Management. }Managing sensitive information, such as API keys or credentials, is more complex in serverless architectures. The stateless nature of serverless functions means that secrets cannot be stored in the same way as in stateful servers. This can introduce security vulnerabilities if handled incorrectly, making secrets management a critical and unique issue in serverless computing. Lenarduzzi et al.~\cite{Lenarduzzi} identified management of secrets as a cause of TD, complementing the findings of our study.

\textbf{Testing Locally. }Testing serverless functions locally can introduce significant challenges due to the tightly coupled nature of these functions with cloud services. Simulating the cloud environment locally can be challenging and lead to discrepancies between local test results and actual behavior in production~\cite{De}. This issue is specific to serverless due to its integration with managed cloud services, which are not easily replicated in a local testing environment. Lenarduzzi et al.~\cite{Lenarduzzi} identified testing serverless functions as a difficult task that can accrue TD if not done properly and on time.

Baldini et al.~\cite{baldini} identified issues related to cold start, scaling, accessing logs, and minimizing cost, whereas Wen et al.~\cite{wen2021empirical} identified issues related to cold start and accessing logs. However, our analysis focused on the TD perspective, while theirs concentrated on identifying common issues in serverless. Our work demonstrates that, if unresolved, the aforementioned issues gradually accrue TD. Through expert interviews, Lenarduzzi et al.~\cite{Lenarduzzi} determined that testing locally, secret management, and accessing logs are activities that generate TD. Our research is complementary, offering empirical support from a broader perspective (SO users).

\subsection{Availability of Solutions to Serverless TD }
\label{subsec:discussionC}

The pattern in the number of answers and accepted answers for serverless-related questions shows that TD questions are more likely to remain unanswered than non-TD questions. Out of 29,212 TD-related questions, 21\% remain unanswered, and 40\% have no accepted answers. To compare, out of 49,655 non-TD questions, 19\% remain unanswered, and 40\% have no accepted answers.

Even though there may be several reasons behind unanswered questions~\cite{Asaduzzaman, Saha}, the high percentage of unanswered TD questions signifies limited knowledge and expertise in this area. It raises questions about conducting further research, involving the community, and providing better resources to help developers handle TD in serverless systems. As serverless computing evolves, the community's ability to offer end-to-end solutions to such TD-related problems will be determinant to the continued success of serverless computing.

\subsection{Code Debt in Serverless Computing }
\label{subsec:discussionE}

For RQ3, we categorized the serverless TD questions into 11 TD types. We identified that \textit{code debt} is the most common type of TD. Code debt refers to poorly written code that deviates from best practices~\cite{Li}. However, this might be because serverless-related tags are often used along with other tags. For the study, we collected all SO posts that had one of the selected relevant tags. Therefore, the selected posts may have multiple tags, including those that do not belong to serverless.

\begin{table}[!htbp]
\centering
\caption{Distribution of Serverless-Related Tags}
\begin{tabular}{L{2.8cm} R{1.5cm} R{1cm} R{1.5cm}}
\toprule
\textbf{Tag} & \textbf{Count (Anywhere)} & \textbf{Count (Alone)} & \textbf{Percentage (\%)} \\
\midrule
aws-lambda & 13211 & 125 & 0.95 \\
azure-functions & 6685 & 178 & 2.66 \\
google-cloud-functions & 5467 & 54 & 0.99 \\
serverless & 1786 & 9 & 0.50 \\
serverless-framework & 1570 & 51 & 3.25 \\
vercel & 1436 & 29 & 2.02 \\
aws-serverless & 616 & 2 & 0.32 \\
aws-sam & 526 & 9 & 1.71 \\
aws-sam-cli & 237 & 3 & 1.27 \\
localstack & 156 & 3 & 1.92 \\
serverless-architecture & 115 & 3 & 2.61 \\
openwhisk & 60 & 5 & 8.33 \\
serverless-offline & 54 & 0 & 0.00 \\
faas & 25 & 0 & 0.00 \\
serverless-plugins & 23 & 0 & 0.00 \\
\bottomrule
\end{tabular}
\label{table:d1}
\end{table}

Table~\ref{table:d1} shows a breakdown of the number of questions per tag with that tag as the only tag in the posts (count (alone)) and the number of questions that included that tag in the post, but also included other tags (count (anywhere)), and the percentage of tags count (alone) of the count (anywhere). If we consider the tag ``aws-lambda,'' only 0.95\% of all questions have it as the only tag. This may explain why the percentage of \textit{code debt} in serverless TD questions is higher than in other TD types. Therefore, it is evident that SO users discuss serverless-related TD questions using many other tags besides those specifically related to serverless.

\subsection{Evolution of TD in Serverless SO Questions Over the Years }
\label{subsec:discussionD}

\begin{figure}[h]
\centering
    \includegraphics[width=0.5\textwidth]{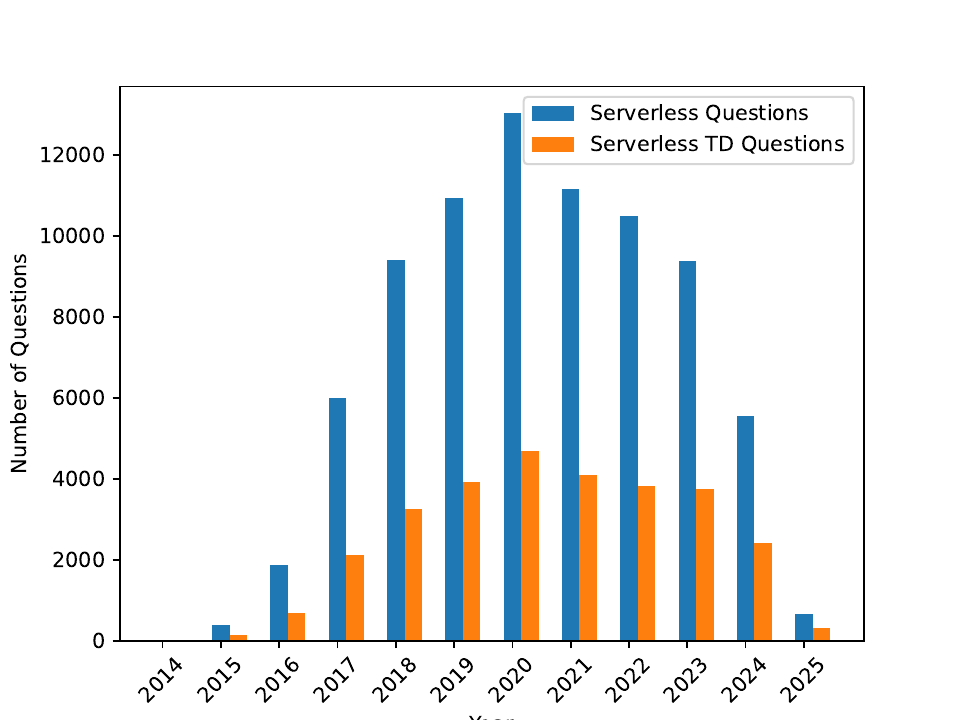}
    \caption{Evolution of Serverless TD Questions Over the Years}
\label{figure:Evolution}
\end{figure}

We used the results of the binary text classifier in RQ2 to investigate the evolution of TD in serverless SO questions from 2014 to 2025. Figure~\ref{figure:Evolution} shows the number of total serverless questions and TD questions over the years. The year 2020 had the highest number of serverless questions on SO, while 2014 had the fewest, since that was the year serverless technologies were introduced. Similarly, 2020 has the highest number of serverless TD questions asked on SO. The overall trend for serverless SO questions shows a gradual increase until 2020, followed by a decrease.

\section{Implications}
\label{sec:implications}

\textbf{For Researchers. }Our study points out various serverless-specific challenges that offer a rich area for exploration. Common types of TD present in serverless architecture, such as the accumulation of TD in code, design, and security, can be studied in greater detail by examining serverless applications to understand how they appear in serverless code. In addition, further research can be conducted to explore how different types of TD evolve over time, how they impact application performance and code maintainability, and how effectively different solutions resolve them. Based on the findings of this research, researchers can develop more robust frameworks and guidelines for addressing TD in serverless computing.

\textbf{For Practitioners. }Practitioners can utilize our findings to understand how TD in serverless computing is being tackled. By identifying the most commonly encountered types and subtypes of TD and their solutions, our research can serve as a reference for developers to identify similar problems in their applications and fill those gaps. Senior developers can leverage these insights to refine their strategies for managing TD, and junior developers can use this experience to better deal with the complexities of serverless computing in production. Our findings can help practitioners learn from shared mistakes and make informed decisions when employing serverless technology.

\section{Threats to Validity}
\label{sec:threatstovalidity}

We identify the threats to the validity of our study following the classification by Runeson et al.~\cite{Runeson}.

\textbf{Internal validity} is the degree to which the ``treatment,'' not other variables, is responsible for the observed results. The classification methodology used to detect serverless TD questions poses the greatest threat to the internal validity of our study. The basis for training the classification model was the BERT model~\cite{He}, achieving an F1-score of 0.86. It's important to note that potential inaccuracies in the dataset might influence results, leading to false positives and negatives. However, for automatic TD classification, we trained multiple models and selected the one with the highest accuracy. In addition to BERT, we trained four models and selected the one with the highest F1-score. Also, we manually classified a statistically significant sample of questions as TD-related or not. We then evaluated the model on the sample, achieving an F1-score of 0.82. Although LLMs like ChatGPT\footnote{https://chatgpt.com/} have made SO less popular, we still consider it an invaluable resource for software engineering research. 62\% of the code produced by LLMs contains API misuse, according to a recent study by Zhong et al.~\cite{zhong2024can} that compared LLMs and SO on coding questions. According to another study by Kabir et al.~\cite{kabir2024stack}, 72\% of ChatGPT responses were verbose, and 52\% contained inaccurate information. Therefore, we regard SO to have higher reliability and precision in answering SO questions, as demonstrated by recent studies, even though we acknowledge developers' use of LLMs like ChatGPT when developing software.

\textbf{External validity} regards the generalizability of our study. We conducted the study using SO questions. However, there are other platforms such as GitHub\footnote{https://github.com/} and SE Stack Exchange\footnote{https://softwareengineering.stackexchange.com/} which can be used for the same purpose. Therefore, we cannot generalize our results. However, as mentioned in the study by Kozanidis et al.~\cite{Kozanidis}, with over 24 million questions and 20 million registered users, SO is the most popular programming Q\&A website among the well-known group of Q\&A websites. We consider SO the best data source for developers to discuss issues related to TD and serverless software development, given its size and popularity.

\textbf{Construct validity} pertains to the extent to which our operational measures are appropriately aligned and capable of addressing our RQs. We extracted the data using the SO tags that we identified. There can be errors when SO users add tags to the questions. To mitigate this threat, we manually validated the selected sample in Section \ref{subsec:methodB} to ensure the questions were related to the mentioned tags.

\textbf{Conclusion validity} relates to how our treatments and outcomes relate to one another. Using a set of tags we determined to be relevant to serverless, we extracted questions about serverless. The results of our study will be affected by the possibility that additional tags are associated with serverless than the eight we examined. Furthermore, we obtained F1, precision, and recall scores of 0.86 for our classification model. This implies that there may have been false positives and negatives in our model, which could have affected our outcomes. Another threat is the manual categorization of the TD questions. Manual categorization may introduce bias and subjective judgment for the analysis. To reduce bias, we asked a graduate student to independently classify a sample of questions, and the interrater agreement was 0.94 (Cohen's Kappa).

\section{Conclusions and Future Work}
\label{sec:conclusion}

In this study, we investigated TD in serverless by analyzing 78,867 SO questions. We first classified serverless-related SO questions as TD- or non-TD-related by training a binary text classifier. Then, we manually analyzed a statistically significant sample of serverless TD questions to identify the types, subtypes of TD, and their solutions. We identified that a significant amount (37\%) of the serverless questions on SO are TD-related, and SO users discussed TD explicitly and used similar words. Manual analysis of the sample of questions revealed 10 types and 49 subtypes of TD, with \textit{code debt} being the most common type, followed by \textit{design} and \textit{versioning debt}, while \textit{infrastructure debt} was the least common type of TD. Our study highlighted issues specific to serverless that complemented existing research. We also found that 21\% of TD questions were left unanswered and 40\% without accepted answers. For the questions with solutions, SO users suggested different solution types, such as code modifications, provided conceptual understandings, and best practices.

For future work, we plan to extend our study to other related platforms, e.g., GitHub issues, and Stack Exchange. Investigating TD across multiple platforms will provide additional insights that may not be captured by using SO alone. In addition, we plan to refine the classification model to get a higher performance, thereby increasing the reliability of the model's output.

\begin{acks}
This study is partly supported by the Natural Sciences and Engineering Research Council of Canada (NSERC), RGPIN-2021-04232 and DGECR-2021-00283, and an NSERC Collaborative Research and Training Experience (CREATE) grant on Software Analytics at the University of Saskatchewan.
\end{acks}

\balance
\bibliographystyle{ACM-Reference-Format}
\bibliography{bibliography}

@article{Cinque,
  title={Real-time faas: serverless computing for industry 4.0},
  author={Cinque, Marcello},
  journal={Service Oriented Computing and Applications},
  volume={17},
  number={2},
  pages={73--75},
  year={2023},
  publisher={Springer}
}

@article{Leitner,
title = {A mixed-method empirical study of Function-as-a-Service software development in industrial practice},
journal = {Journal of Systems and Software},
volume = {149},
pages = {340-359},
year = {2019},
issn = {0164-1212},
doi = {https://doi.org/10.1016/j.jss.2018.12.013},
url = {https://www.sciencedirect.com/science/article/pii/S0164121218302735},
author = {Philipp Leitner and Erik Wittern and Josef Spillner and Waldemar Hummer}}

@article{baldini,
  title={Serverless computing: Current trends and open problems},
  author={Baldini, Ioana and Castro, Paul and Chang, Kerry and Cheng, Perry and Fink, Stephen and Ishakian, Vatche and Mitchell, Nick and Muthusamy, Vinod and Rabbah, Rodric and Slominski, Aleksander and others},
  journal={Research advances in cloud computing},
  pages={1--20},
  year={2017},
  publisher={Springer}
}

@inproceedings{Brown,
author = {Brown, Nanette and Cai, Yuanfang and Guo, Yuepu and Kazman, Rick and Kim, Miryung and Kruchten, Philippe and Lim, Erin and MacCormack, Alan and Nord, Robert and Ozkaya, Ipek and Sangwan, Raghvinder and Seaman, Carolyn and Sullivan, Kevin and Zazworka, Nico},
title = {Managing Technical Debt in Software-Reliant Systems},
year = {2010},
isbn = {9781450304276},
publisher = {Association for Computing Machinery},
address = {New York, NY, USA},
url = {https://doi.org/10.1145/1882362.1882373},
doi = {10.1145/1882362.1882373},
booktitle = {Proceedings of the FSE/SDP Workshop on Future of Software Engineering Research},
pages = {47–52},
numpages = {6},
series = {FoSER '10}
}

@article{Li,
  title={A systematic mapping study on technical debt and its management},
  author={Li, Zengyang and Avgeriou, Paris and Liang, Peng},
  journal={Journal of Systems and Software},
  volume={101},
  pages={193--220},
  year={2015},
  publisher={Elsevier}
}

@article{Gama,
author = {Gama, Eliakim and Paixao, Matheus and Freire, Sávio and Cortés, Mariela},
year = {2019},
month = {10},
pages = {228-233},
title = {Technical Debt's State of Practice on Stack Overflow: a Preliminary Study},
isbn = {978-1-4503-7282-4},
journal = {SBQS'19: Proceedings of the XVIII Brazilian Symposium on Software Quality},
doi = {10.1145/3364641.3364668}
}

@article{Costa,
author = {Costa, Diego and Cort\'{e}s, Mariela and Gama, Eliakim},
title = {On the Relation between Technical Debt Indicators and Quality Criteria in Stack Overflow Discussions},
year = {2021},
isbn = {9781450390613},
publisher = {Association for Computing Machinery},
address = {New York, NY, USA},
url = {https://doi.org/10.1145/3474624.3474648},
doi = {10.1145/3474624.3474648},
booktitle = {Proceedings of the XXXV Brazilian Symposium on Software Engineering},
pages = {432–441},
numpages = {10},
location = {Joinville, Brazil},
series = {SBES '21}
}

@article{Kozanidis,
author = {Kozanidis, Nicholas and Verdecchia, Roberto and Guzman, Emitza},
title = {Asking about Technical Debt: Characteristics and Automatic Identification of Technical Debt Questions on Stack Overflow},
year = {2022},
isbn = {9781450394277},
publisher = {Association for Computing Machinery},
address = {New York, NY, USA},
url = {https://doi.org/10.1145/3544902.3546245},
doi = {10.1145/3544902.3546245},
booktitle = {Proceedings of the 16th ACM / IEEE International Symposium on Empirical Software Engineering and Measurement},
pages = {45–56},
numpages = {12},
location = {Helsinki, Finland},
series = {ESEM '22}
}

@article{Gama2,
author = {Gama, Eliakim and Freire, S\'{a}vio and Mendon\c{c}a, Manoel and Sp\'{\i}nola, Rodrigo O. and Paixao, Matheus and Cort\'{e}s, Mariela I.},
title = {Using Stack Overflow to Assess Technical Debt Identification on Software Projects},
year = {2020},
isbn = {9781450387538},
publisher = {Association for Computing Machinery},
address = {New York, NY, USA},
url = {https://doi.org/10.1145/3422392.3422429},
doi = {10.1145/3422392.3422429},
booktitle = {Proceedings of the XXXIV Brazilian Symposium on Software Engineering},
pages = {730–739},
numpages = {10},
location = {Natal, Brazil},
series = {SBES '20}
}

@inproceedings{Edbert,
  title={Exploring Technical Debt in Security Questions on Stack Overflow},
  author={Edbert, Joshua Aldrich and Oishwee, Sahrima Jannat and Karmakar, Shubhashis and Codabux, Zadia and Verdecchia, Roberto},
  booktitle={2023 ACM/IEEE International Symposium on Empirical Software Engineering and Measurement (ESEM)},
  pages={1--12},
  year={2023},
  organization={IEEE}
}

@article{Yang,
  title={What security questions do developers ask? a large-scale study of stack overflow posts},
  author={Yang, Xin-Li and Lo, David and Xia, Xin and Wan, Zhi-Yuan and Sun, Jian-Ling},
  journal={Journal of Computer Science and Technology},
  volume={31},
  pages={910--924},
  year={2016},
  publisher={Springer}
}

@article{Titan,
author={Tian, Fangchao and Liang, Peng and Babar, Muhammad Ali},
booktitle={2019 IEEE International Conference on Software Architecture (ICSA)}, 
title={How Developers Discuss Architecture Smells? An Exploratory Study on Stack Overflow}, 
year={2019},
volume={},
number={},
pages={91-100},
doi={10.1109/ICSA.2019.00018}
}

@article{Taibi,
author={Taibi, Davide and Kehoe, Ben and Poccia, Danilo},
booktitle={2022 IEEE International Conference on Service-Oriented System Engineering (SOSE)}, 
title={Serverless: From Bad Practices to Good Solutions}, 
year={2022},
volume={},
number={},
pages={85-92},
doi={10.1109/SOSE55356.2022.00016}
}

@article{Lenarduzzi,
author={Lenarduzzi, Valentina and Daly, Jeremy and Martini, Antonio and Panichella, Sebastiano and Tamburri, Damian Andrew},
journal={IEEE Software}, 
title={Toward a Technical Debt Conceptualization for Serverless Computing}, 
year={2021},
volume={38},
number={1},
pages={40-47},
doi={10.1109/MS.2020.3030786}
}

@article{Toledo,
author = {Toledo, Saulo and Martini, Antonio and Sjøberg, Dag},
year = {2021},
month = {04},
pages = {110968},
title = {Identifying architectural technical debt, principal, and interest in microservices: A multiple-case study},
volume = {177},
journal = {Journal of Systems and Software},
doi = {10.1016/j.jss.2021.110968}
}

@article{Pigazzini,
author = {Pigazzini, Ilaria and Fontana, Francesca Arcelli and Lenarduzzi, Valentina and Taibi, Davide},
title = {Towards Microservice Smells Detection},
year = {2020},
isbn = {9781450379601},
publisher = {Association for Computing Machinery},
address = {New York, NY, USA},
url = {https://doi.org/10.1145/3387906.3388625},
doi = {10.1145/3387906.3388625},
booktitle = {Proceedings of the 3rd International Conference on Technical Debt},
pages = {92–97},
numpages = {6},
location = {Seoul, Republic of Korea},
series = {TechDebt '20}
}

@book{Wohlin,
  title={Experimentation in software engineering},
  author={Wohlin, Claes and Runeson, Per and H{\"o}st, Martin and Ohlsson, Magnus C and Regnell, Bj{\"o}rn and Wessl{\'e}n, Anders and others},
  volume={236},
  year={2012},
  publisher={Springer}
}

@article{Avgeriou,
  title={Managing technical debt in software engineering (dagstuhl seminar 16162)},
  author={Avgeriou, Paris and Kruchten, Philippe and Ozkaya, Ipek and Seaman, Carolyn},
  journal={Dagstuhl reports},
  volume={6},
  number={4},
  pages={110--138},
  year={2016},
  publisher={Schloss Dagstuhl--Leibniz-Zentrum f{\"u}r Informatik}
}

@article{Breiman,
  title={Random forests},
  author={Breiman, Leo},
  journal={Machine learning},
  volume={45},
  pages={5--32},
  year={2001},
  publisher={Springer}
}

@article{He,
  title={Deberta: Decoding-enhanced bert with disentangled attention},
  author={He, Pengcheng and Liu, Xiaodong and Gao, Jianfeng and Chen, Weizhu},
  journal={arXiv preprint arXiv:2006.03654},
  year={2020}
}

@article{Devlin,
  title={Bert: Pre-training of deep bidirectional transformers for language understanding},
  author={Devlin, Jacob},
  journal={arXiv preprint arXiv:1810.04805},
  year={2018}
}

@inproceedings{Van,
  title={A SPEC RG cloud group's vision on the performance challenges of FaaS cloud architectures},
  author={Van Eyk, Erwin and Iosup, Alexandru and Abad, Cristina L and Grohmann, Johannes and Eismann, Simon},
  booktitle={Companion of the 2018 acm/spec international conference on performance engineering},
  pages={21--24},
  year={2018}
}

@article{Tran,
  title={Concurrent service auto-scaling for Knative resource quota-based serverless system},
  author={Tran, Minh-Ngoc and Kim, YoungHan},
  journal={Future Generation Computer Systems},
  year={2024},
  publisher={Elsevier}
}

@inproceedings{Lloyd,
  title={Serverless computing: An investigation of factors influencing microservice performance},
  author={Lloyd, Wes and Ramesh, Shruti and Chinthalapati, Swetha and Ly, Lan and Pallickara, Shrideep},
  booktitle={2018 IEEE international conference on cloud engineering (IC2E)},
  pages={159--169},
  year={2018},
  organization={IEEE}
}

@inproceedings{Somma,
  title={When less is more: Core-restricted container provisioning for serverless computing},
  author={Somma, Gaetano and Ayimba, Constantine and Casari, Paolo and Romano, Simon Pietro and Mancuso, Vincenzo},
  booktitle={IEEE INFOCOM 2020-IEEE Conference on Computer Communications Workshops (INFOCOM WKSHPS)},
  pages={1153--1159},
  year={2020},
  organization={IEEE}
}

@article{Yu,
  title={Serverless Cold Start Performance Optimization Based on Multi-Request Processing and Adaptive Hierarchical Scaling},
  author={Yu, Liu and Fu, Li and Chenhao, Sun},
  journal={IEEE Access},
  year={2024},
  publisher={IEEE}
}

@inproceedings{Agarwal,
  title={A reinforcement learning approach to reduce serverless function cold start frequency},
  author={Agarwal, Siddharth and Rodriguez, Maria A and Buyya, Rajkumar},
  booktitle={2021 IEEE/ACM 21st International Symposium on Cluster, Cloud and Internet Computing (CCGrid)},
  pages={797--803},
  year={2021},
  organization={IEEE}
}

@article{Ghorbian,
  title={A survey on the cold start latency approaches in serverless computing: an optimization-based perspective},
  author={Ghorbian, Mohsen and Ghobaei-Arani, Mostafa},
  journal={Computing},
  pages={1--55},
  year={2024},
  publisher={Springer}
}

@inproceedings{Vahidinia,
  title={Cold start in serverless computing: Current trends and mitigation strategies},
  author={Vahidinia, Parichehr and Farahani, Bahar and Aliee, Fereidoon Shams},
  booktitle={2020 International Conference on Omni-layer Intelligent Systems (COINS)},
  pages={1--7},
  year={2020},
  organization={IEEE}
}

@inproceedings{Verma,
  title={A Review: Cold Start Latency in Serverless Computing},
  author={Verma, Parikshit and Goel, Paurav and Rani, Neetu},
  booktitle={2024 Sixth International Conference on Computational Intelligence and Communication Technologies (CCICT)},
  pages={141--148},
  year={2024},
  organization={IEEE}
}

@article{Ebrahimi,
  title={Cold start latency mitigation mechanisms in serverless computing: taxonomy, review, and future directions},
  author={Ebrahimi, Ana and Ghobaei-Arani, Mostafa and Saboohi, Hadi},
  journal={Journal of Systems Architecture},
  pages={103115},
  year={2024},
  publisher={Elsevier}
}

@article{Karmazadeh,
  title={Reducing cold start delay in serverless computing using lightweight virtual machines},
  author={Karamzadeh, Amirmohammad and Shameli-Sendi, Alireza and Dagenais, Michel},
  journal={Journal of Network and Computer Applications},
  pages={104030},
  year={2024},
  publisher={Elsevier}
}

@article{Manner,
  title={Troubleshooting serverless functions: a combined monitoring and debugging approach},
  author={Manner, Johannes and Kolb, Stefan and Wirtz, Guido},
  journal={SICS Software-Intensive Cyber-Physical Systems},
  volume={34},
  pages={99--104},
  year={2019},
  publisher={Springer}
}

@inproceedings{Sedefo,
  title={Cost minimization for deploying serverless functions},
  author={Sedefo{\u{g}}lu, {\"O}zg{\"u}r and S{\"o}zer, Hasan},
  booktitle={Proceedings of the 36th Annual ACM Symposium on Applied Computing},
  pages={83--85},
  year={2021}
}

@article{Sarroca,
  title={Mlless: Achieving cost efficiency in serverless machine learning training},
  author={Sarroca, Pablo Gimeno and S{\'a}nchez-Artigas, Marc},
  journal={Journal of Parallel and Distributed Computing},
  volume={183},
  pages={104764},
  year={2024},
  publisher={Elsevier}
}

@article{De,
  title={The Impact of Software Testing on Serverless Applications},
  author={De Silva, Dilshan and Hewawasam, Lakindu},
  journal={IEEE Access},
  year={2024},
  publisher={IEEE}
}

@inproceedings{Asaduzzaman,
  title={Answering questions about unanswered questions of stack overflow},
  author={Asaduzzaman, Muhammad and Mashiyat, Ahmed Shah and Roy, Chanchal K and Schneider, Kevin A},
  booktitle={2013 10th Working Conference on Mining Software Repositories (MSR)},
  pages={97--100},
  year={2013},
  organization={IEEE}
}

@inproceedings{Saha,
  title={Toward understanding the causes of unanswered questions in software information sites: a case study of stack overflow},
  author={Saha, Ripon K and Saha, Avigit K and Perry, Dewayne E},
  booktitle={Proceedings of the 2013 9th Joint Meeting on Foundations of Software Engineering},
  pages={663--666},
  year={2013}
}

@inproceedings{Pina,
  title={Technical debt prioritization: Taxonomy, methods results, and practical characteristics},
  author={Pina, Diogo and Goldman, Alfredo and Tonin, Graziela},
  booktitle={2021 47th Euromicro Conference on Software Engineering and Advanced Applications (SEAA)},
  pages={206--213},
  year={2021},
  organization={IEEE}
}

@inproceedings{Alves,
  title={Towards an ontology of terms on technical debt},
  author={Alves, Nicolli SR and Ribeiro, Leilane F and Caires, Vivyane and Mendes, Thiago S and Sp{\'\i}nola, Rodrigo O},
  booktitle={2014 Sixth International Workshop on Managing Technical Debt},
  pages={1--7},
  year={2014},
  organization={IEEE}
}

@article{AlvesMappingStudy,
  title={Identification and management of technical debt: A systematic mapping study},
  author={Alves, Nicolli SR and Mendes, Thiago S and De Mendon{\c{c}}a, Manoel G and Sp{\'\i}nola, Rodrigo O and Shull, Forrest and Seaman, Carolyn},
  journal={Information and Software Technology},
  volume={70},
  pages={100--121},
  year={2016},
  publisher={Elsevier}
}

@article{Rios,
  title={A tertiary study on technical debt: Types, management strategies, research trends, and base information for practitioners},
  author={Rios, Nicolli and de Mendon{\c{c}}a Neto, Manoel Gomes and Sp{\'\i}nola, Rodrigo Oliveira},
  journal={Information and Software Technology},
  volume={102},
  pages={117--145},
  year={2018},
  publisher={Elsevier}
}

@article{Runeson,
  title={Guidelines for conducting and reporting case study research in software engineering},
  author={Runeson, Per and H{\"o}st, Martin},
  journal={Empirical software engineering},
  volume={14},
  pages={131--164},
  year={2009},
  publisher={Springer}
}

@inproceedings{wen2021empirical,
  title={An empirical study on challenges of application development in serverless computing},
  author={Wen, Jinfeng and Chen, Zhenpeng and Liu, Yi and Lou, Yiling and Ma, Yun and Huang, Gang and Jin, Xin and Liu, Xuanzhe},
  booktitle={Proceedings of the 29th ACM joint meeting on European software engineering conference and symposium on the foundations of software engineering},
  pages={416--428},
  year={2021}
}

@article{li2022serverless,
  title={Serverless computing: state-of-the-art, challenges and opportunities},
  author={Li, Yongkang and Lin, Yanying and Wang, Yang and Ye, Kejiang and Xu, Chengzhong},
  journal={IEEE Transactions on Services Computing},
  volume={16},
  number={2},
  pages={1522--1539},
  year={2022},
  publisher={IEEE}
}

@article{tanzil2025systematic,
  title={A systematic mapping study of crowd knowledge enhanced software engineering research using Stack Overflow},
  author={Tanzil, Minaoar Hossain and Chowdhury, Shaiful and Modaberi, Somayeh and Uddin, Gias and Hemmati, Hadi},
  journal={Journal of Systems and Software},
  pages={112405},
  year={2025},
  publisher={Elsevier}
}

@inproceedings{zhong2024can,
  title={Can llm replace stack overflow? a study on robustness and reliability of large language model code generation},
  author={Zhong, Li and Wang, Zilong},
  booktitle={Proceedings of the AAAI conference on artificial intelligence},
  volume={38},
  number={19},
  pages={21841--21849},
  year={2024}
}

@inproceedings{kabir2024stack,
  title={Is stack overflow obsolete? an empirical study of the characteristics of chatgpt answers to stack overflow questions},
  author={Kabir, Samia and Udo-Imeh, David N and Kou, Bonan and Zhang, Tianyi},
  booktitle={Proceedings of the 2024 CHI Conference on Human Factors in Computing Systems},
  pages={1--17},
  year={2024}
}

@article{blair2015reflexive,
  title={A reflexive exploration of two qualitative data coding techniques},
  author={Blair, Erik},
  journal={Journal of Methods and Measurement in the Social Sciences},
  volume={6},
  number={1},
  pages={14--29},
  year={2015}
}

@article{cohen1960coefficient,
  title={A coefficient of agreement for nominal scales},
  author={Cohen, Jacob},
  journal={Educational and psychological measurement},
  volume={20},
  number={1},
  pages={37--46},
  year={1960},
  publisher={Sage Publications Sage CA: Thousand Oaks, CA}
}

@article{mayring2014qualitative,
  title={Qualitative content analysis: theoretical foundation, basic procedures and software solution},
  author={Mayring, Philipp},
  year={2014},
  publisher={AUT}
}
\end{document}